\documentclass[aps,prb,twocolumn,floatfix,nolongbibliography]{revtex4-2}

\usepackage{amsmath}

\usepackage{amssymb}
\usepackage{graphicx}
\usepackage{bm}
\usepackage{color}
\usepackage{times}
\begin{document}

\title{Analytic Solution of $n$-dimensional Su-Schrieffer–Heeger Model}

\author{Feng Liu$^{1,2,3}$}
\email{Liufeng@nbu.edu.cn}

\affiliation{$^{1}$Institute of High Pressure Physics, Ningbo
  University, Ningbo, 315-211, China}

\affiliation{$^{2}$School of Physical Science and Technology, Ningbo
  University, Ningbo, 315-211, China}

\affiliation{$^{3}$Department of Nanotechnology for Sustainable Energy, School of Science and Technology, Kwansei Gakuin University, Gakuen 2-1, Sanda 669-1337, Japan}

\begin{abstract}
The Su-Schrieffer-Heeger (SSH) model is fundamental in topological insulators and relevant to understanding higher-order topological phases. 
This study explores the relationship between the $n$-dimensional SSH model and its $(n-1)$-dimensional counterpart, identifying a hierarchical structure in the Hamiltonian that allows us to solve an arbitrary $n$-dimensional SSH model analytically. 
By generalizing the bulk-edge correspondence principle to arbitrary dimensions in a higher-order fashion using the vectored Zak phase, we reveal a type of topological insulator called hierarchical topological insulators. 
In this hierarchical topological insulator, there exist intermediate-order topological interfacial states that are protected by subsymmetries and energy bands topology in a partial Brillouin zone. 
Furthermore, we compare the $n$-dimensional SSH model with the Benalcazar-Bernevig-Hughes (BBH) model, another essential model in higher-order topological phases similar to the two-dimensional SSH model with an extra flux of $\pi$ in each plaque. We find that the BBH model is another example of hierarchical topological insulators.    
\end{abstract}

\maketitle
\section{Introduction}
Topology has introduced a new perspective for classifying crystalline systems based on the geometric properties of their energy bands in momentum space, as opposed to their band gap sizes~\cite{Hasan2010, Qi2011, Bansil2016}. 
This new theory, called band topology theory, has revolutionized our understanding of solid-state physics.
When topological invariants contrast, it is possible to differentiate between crystalline systems with seemingly identical band structures through topological interfacial states.
At the core of the band topology theory lies the bulk-edge correspondence, which establishes a link between the bulk topological invariant of crystalline systems and the emergence of robust interfacial states between topologically distinct finite samples~\cite{Hatsugai1993, Liang2006, Hwang2019, Bouhon2019, Wang2021}. 
These topological interfacial states are not affected by perturbations of amplitudes smaller than band gaps and large-amplitude perturbations that respect specific symmetries, such as time-reversal symmetry. 
This property makes them promising for potential transformative applications, such as topological quantum computation and scattering-free wave transport.

The bulk-edge correspondence has recently been extended to a higher-order fashion, where interfacial states of $(n-d)$ dimensions appear in $n$-dimensional ($n$D) systems for $d>1$~\cite{Benalcazar2017a, Benalcazar2017B, Song2017, Langbehn2017, Khalaf2018, Geier2018, Ezawa2018, Schindler2018, Schindler2018a, Xie2018, Ezawa2018a, Wang2018a, Zhu2018, Queiroz2019, Yan2019, Zhang2019b, Luo2019, Xie2019, Liu2019, Park2019, Yue2019, Okugawa2019, LiuBing2019, Liu2019a, Zhang2019d, Trifunovic2019, Peng2020, Cerjan2020, Chen2020, Hua2020, Hu2020, Huang2020, Choi2020, Zhang2020, Nag2021,Aggarwal2021, Ghosh2021, Khalaf2021, Wang2021a, Zhang2021a,Benalcazar2022, Tan2022, Lei2022, Jia2022,Ghosh2022A,Ghosh2022B,Saha2023}. 
This means that topological corner states emerge owing to the bulk-corner correspondence. These corner states usually accompany filling-anomaly-induced fractional charges and have potential applications in fields such as laser cavity and quantum computation~\cite{Hararieaar2018, Benalcazar2019, Wu2020, Watanabe2021, Jung2021, Zhang2020c, Zhang2020d, Pahomi2020, Takahashi2021, Li2021b, Pan2022, Liu2021a,Tang2023, He2023}. 
Specifically, the emergence of topological corner states has opened up new possibilities for designing robust and efficient devices for these applications. For example, corner states can be used as topological waveguides in laser cavities to achieve lasing with higher efficiency and greater stability. Similarly, they can be utilized in topological quantum computation to improve scalability and fault tolerance. 
The discovery of topological corner states has broadened the scope of topological materials research and opened new avenues for technological advancement. 
Over the past few years, there has been significant growth in higher-order band topology, as demonstrated by several models, including the 2D Su-Schrieffer–Heeger (SSH) model~\cite{Liu2017}, the Benalcazar-Bernevig-Hughes (BBH) model~\cite{Benalcazar2017a, Benalcazar2017B}, and the breathing Kagome lattice model~\cite{Ezawa2018}. These models have allowed the experimental observation of higher-order topological states in various artificial crystalline systems~\cite{Peterson2018, Imhof2018, Serra-Garcia2018, Mittal2019, Zhang2019c, He2020, Xue2020, Dutt2020, Qi2020, Xue2020, Bao2019, Zhang2020e}. 
Although these three models share similar alternative hopping textures featured in the 1D SSH model, whether their higher-order topological nature is essentially the same remains elusive. To answer this question, we analytically solved the $n$D SSH model, which helps us to check the higher-order topological properties in detail.  

This study focuses on the SSH model and extends the 1D model to arbitrary $n$ dimensions by identifying a hierarchical structure in the Hamiltonian of the $n$D SSH model. Leveraging this hierarchical structure, we provide an analytical solution for the $n$D SSH model and generalize the bulk-edge correspondence to arbitrary dimensions in a higher-order fashion. For example, we establish the $n$-0 correspondence using the vectored Zak phase, where $n$ denotes the bulk dimension and 0 denotes the dimension of the topological interfacial states. Furthermore, we show that the topological interfacial states in the $n$D SSH model are protected by subsymmetry related to the hierarchical structure rather than a specific bulk symmetry. This hierarchical structure can also be applied to the BBH model, and we give a preliminary comparison of the topological nature between the 2D SSH model and the BBH model.

Compared to the existing literature that generalizes the BBH model to arbitrary dimensions by using the Clifford algebra and analytical solution of corner states~\cite{Luo2023A, Luo2023B}, our work reveals a hierarchical structure of the $n$D SSH model and $n$D BBH model, which enables us to obtain analytical solutions including eigenenergies and wavefunctions of the $n$D SSH model for bulk, edge, and other higher-order topological states. Furthermore, with the aid of the hierarchical structure, we generalize the bulk-edge correspondence to $n-(n-l)$ correspondence, which brings us various topological phases characterized by intermediate-order topological states that reveal a new type of topological insulators called hierarchical topological insulators. We also compare the $n$D SSH model and the $n$D BBH model, we find that they are both hierarchical topological insulators.     

The structure of the remaining parts is organized as follows. In Sec. II, we give the hierarchical structure $n$D SSH Hamiltonian, its analytical solutions, and the corresponding topological invariants. In Sec. III, we generalize the bulk-edge correspondence in the 1D SSH model to the $n$D SSH model in a fashion of higher order. In Sec. IV, we compare the 2D SSH model with the BBH model. Finally, we give conclusions and discussions in Sec. V.

\section{Hierarchical structure of $n$D SSH model}

The unit cell of the higher-dimensional SSH model can be constructed by aggregating its lower-dimensional version, as depicted in Fig.~1, where we label each sublattice in binary order. For example, in the 1D SSH model, we label the two sublattices as $|0\rangle$ and $|1\rangle$, and in the 2D SSH model, we label the four sublattices as $|00\rangle$, $|01\rangle$, $|10\rangle$ and $|11\rangle$, and so on.  Taking the 2D SSH model as an example, the Hamiltonian can be written in terms of the 1D SSH Hamiltonian as 
\begin{equation}
\mathcal{H}_{2D}=I_2\otimes\mathcal{H}_{1D}+\mathcal{H}_{1D}\otimes I_2,
\end{equation}
where $I_2$ is $2\times 2$ unitary matrix, and $\otimes$ denotes the Kronecker product. This process can be performed recursively for 3D, 4D, and eventually for arbitrary $n$D SSH models. 
We have the $2^n \times 2^n$ matrix as 
\begin{equation}
\begin{split}
\mathcal{H}_{nD}=&I_2\otimes\mathcal{H}_{(n-1)D}+H_{1D}\otimes I^{n-1}_{2}\\
                =&I_2(I_2\mathcal{H}_{(n-2)D}+H_{1D}I^{n-2}_2)+H_{1D} I^{n-1}_2\\
                =&\sum^{n-1}_{i=0}I^i_2H_{1D}I^{n-i-1}_2.
\end{split}
\end{equation}

It is noted that robust corner states could appear if all stacking $1D$ SSH models in Eq.~(2) are nontrivial, which is the parameter setting of $\gamma^\prime_i>\gamma_i$ for all $i$s. These corner states are protected by chiral symmetry, as discussed in Sec.~IIIC.  
In a more transparent form, $H_{nD}$ holds a hierarchical structure between $n$ and $n-1$ dimensional SSH models, which is 
\begin{equation}
    H_{nD}=
    \begin{pmatrix}
    H_{(n-1)D} & \rho_i \\
    \rho^*_i & H_{(n-1)D}
    \end{pmatrix},
\end{equation}
where $\rho_i=\gamma_i+\gamma^\prime_i e^{ik_i}$ is the hopping term between the additional dimension. Equations (2) and (3) can be diagonalized recursively by making use of the eigensolution of the 1D SSH model. Therefore, we can obtain the eigenenergy of the $n$D SSH model as
\begin{equation}
    E^{(nD)}=\sum_i s_i |\rho_i|,
\end{equation}
where $s_i$ takes value $\pm 1$, and $i$ run overs all the dimensions. 
From Eq.~(3), we obtain the recursive relation of the eigenfunction between the SSH models $n$ and $n-1$, which is 
\begin{equation}
\begin{split}
|\psi^{(nD)}, s_1s_2\dots s_{n}\rangle=&\frac{1}{\sqrt{2}}(|\psi^{(n-1)D},s_1s_2\dots s_{n-1}\rangle\\&+s_ne^{-i\phi_n}|\psi^{(n-1)D},s_1s_2\dots s_{n-1}\rangle),
\end{split}
\end{equation}
where $s_i$ takes value $\pm 1$ determining the order of energy bands. Expanding Eq.~(5) by the sublattice bases, we obtain
\begin{equation}
    |\psi^{(nD)}\rangle=\frac{1}{(\sqrt{2})^n}\sum^{n,P^n_m}_{m=0,j}
    \mathcal{S}^j_{n,m}[s_i]e^{-i\mathcal{S}^j_{n,m}[\sum\phi_i]}|\mathcal{S}^j_{n,m}[\underbrace{0\dots00}_{n}]\rangle,
\end{equation}
where $\mathcal{S}^1_{n,m}$, $\mathcal{S}^2_{n,m}$, $\cdots$, $\mathcal{S}^{P^n_m}_{n,m}$ are all the permutations of picking $(0 \leq m \leq n)$ numbers from $n$, and $\mathcal{S}^j_{n,m}[f_i]$ is the corresponding permutation in term of $f_i$. For example, $\mathcal{S}^1_{3,2}[s_i]$ is $s_1s_2$, $\mathcal{S}^1_{3,2}[\sum\phi_i]$ is $\phi_1+\phi_2$, and $|\mathcal{S}^1_{3,2}[000]\rangle$ is $|011\rangle$.    

\begin{figure}[t]
\leavevmode
\begin{center}
\leavevmode
\includegraphics[clip=true,width=0.99\columnwidth]{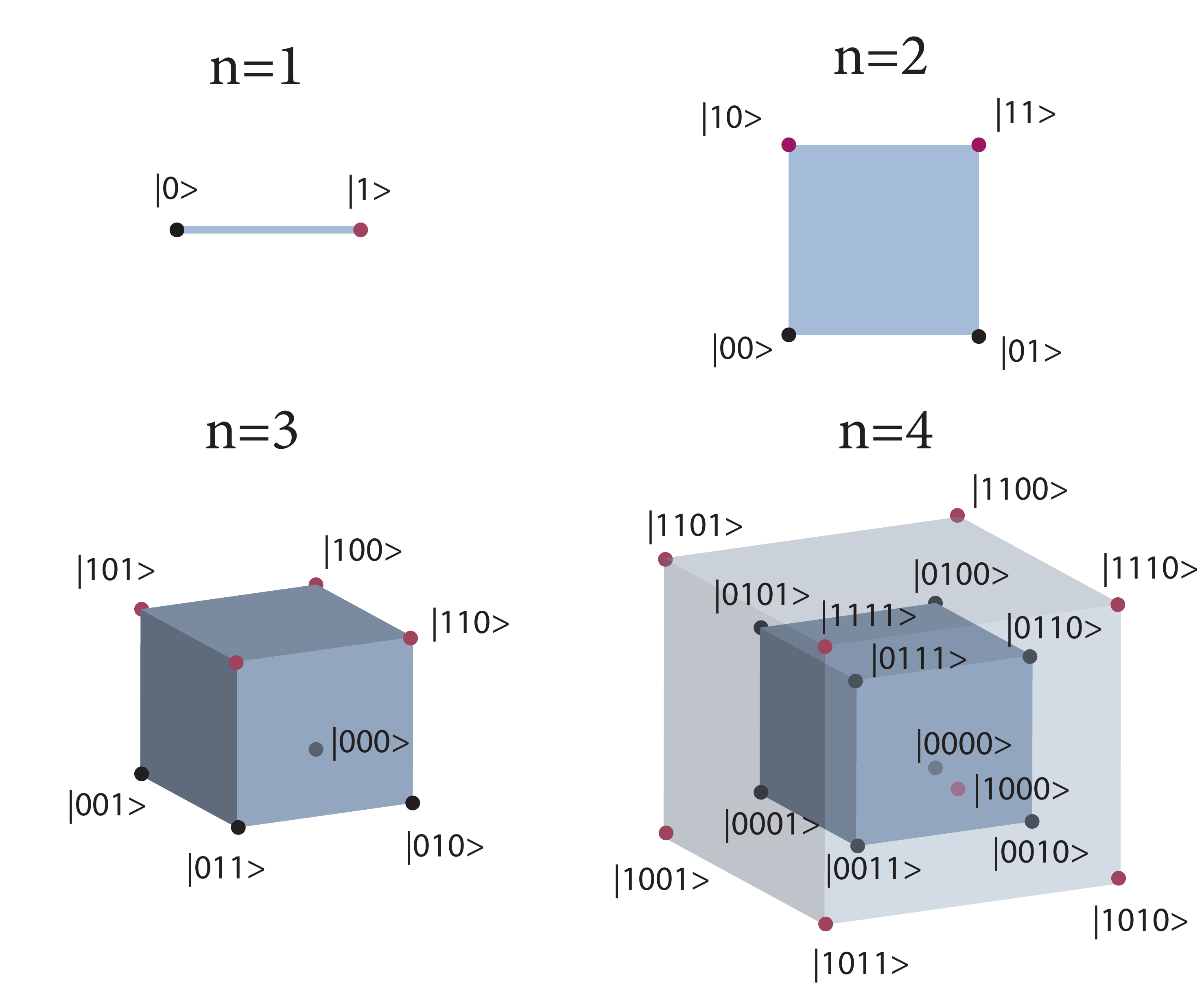}
\end{center}
\caption{Unit cells of the SSH models in 1D, 2D, 3D, and 4D. The black and red atoms indicate the sublattices of the original dimension and the extra dimension, respectively. The sublattices are labeled following a binary order, such as the sublattices of the original dimension starting from $|0\dots\rangle$ and those of the extra dimension starting from $|1\dots\rangle$. The unit cell of the four-dimensional SSH model has a hypercube structure, and a 3D projection is plotted here. }
\end{figure}

\subsection{1D SSH model}
\subsubsection{Bulk spectrum and wave function}
The 1D SSH model Hamiltonian that describes an atomic chain with two types of sublattices and hoppings can be written as~\cite{Heeger1988}
\begin{equation}
\mathcal{H}_{1D}=\sum_N( \gamma a_N^\dagger b_N + \gamma^\prime b^\dagger_{N-1}a_{N})+\text{h.c.},    
\end{equation}
where $N$ represents the number of unit cells, and $\gamma$ and $\gamma^\prime$ are intra-cell and inter-cell hopping amplitudes, respectively. 
By Fourier transformation, Eq.~(7) can be cast into a $2 \times 2$ matrix $\mathcal{H}_{1D}(k)=\text{Re}(\rho(k))\sigma_x-\text{Im}(\rho(k))\sigma_y$ with $\sigma$ the Pauli matrices. 
The eigenvalues and eigenvectors of 1D SSH model is given as
\begin{equation}
    \begin{split}
    E^{1D}_{\pm}&=\pm|\rho|\\
    |\psi^{1D},\pm\rangle&=\frac{1}{\sqrt{2}}
    \begin{pmatrix}
    1 \\
    \pm e^{-i\phi(k)}
    \end{pmatrix}
   \end{split}  
\end{equation}
with $\rho\equiv|\rho|e^{i\phi}=\gamma+\gamma^\prime e^{ik}$ and $k$ the quasi wave-number in crystalline systems. 
We plot the bulk energy spectrum of the 1D SSH model in Fig.~2(a).

\subsubsection{Winding number and Zak phase}
\begin{figure}[t]
\leavevmode
\begin{center}
\leavevmode
\includegraphics[clip=true,width=0.99\columnwidth]{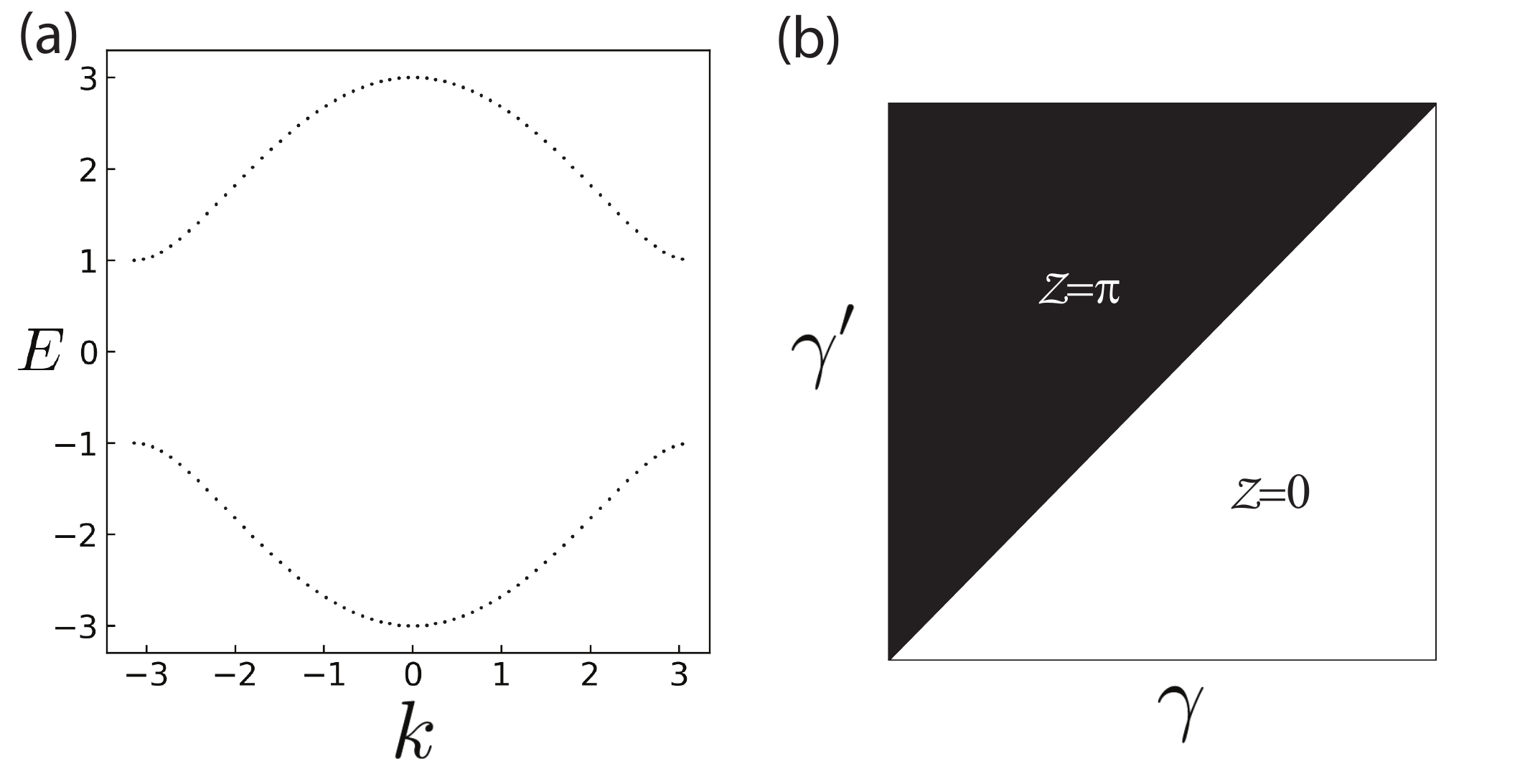}
\end{center}
\caption{(a) Bulk energy spectrum of the 1D SSH model for $|\gamma/\gamma^\prime|=2.0$. 
(b) Topological phase diagram of the 1D SSH model depends on $\gamma$ and $\gamma^\prime$ in terms of the Zak phase.}
\end{figure}

In the 1D SSH model, there exists a chiral symmetry, which is $\mathcal{C}=\sigma_z$.
Effectively, the chiral symmetry can also be regarded as the sublattice symmetry.
Under $\mathcal{C}$, we have $\mathcal{C}^{-1}\mathcal{H}\mathcal{C}=-\mathcal{H}$.
Constrained by this chiral symmetry, the Hamiltonian of the 1D SSH model can always be cast in an off-diagonal form.
We can define a winding number by the off-diagonal term of the 1D SSH Hamiltonian $\rho$. 
The winding number is written as $\nu=\frac{1}{2\pi i}\int_\pi^{\pi} dln(\rho(k))/dk$, which is determined by the winding of $\phi(k)$ around the origin. 
Interestingly, the winding of $\phi(k)$ also coincides with the Zak phase of the wave function $|\psi\rangle$, which is given as $\mathcal{Z}_i=\int_0^{2\pi}dk_i\langle \psi|i\partial k_i|\psi\rangle=\Delta\phi_i/2$.

The winding number and Zak phase describe the topological property of the 1D SSH model. 
Depending on the ratio of $|\gamma/\gamma^\prime|$, the Zak phase can be $\pi$ and 0. 
Figure 2(b) plots the topological phase diagram in terms of $\gamma$ and $\gamma^\prime$. 
For the nontrivial Zak phase $\pi$, topological edge states appear. 

\subsection{2D SSH model}

\subsubsection{Bulk spectrum and wave function}
Using Eqs.~(5) to (8), we are ready to discuss the specific case like $n=2$. 
The bulk energy spectrum of the 2D SSH model is given by
\begin{equation}
E^{(2D)}=s_1|\rho_x|+s_2|\rho_y|,    
\end{equation}
where $s_1$ and $s_2$ take values of $\pm 1$.  The lowest energy band has $s_1=-1$ and $s_2=-1$, and its corresponding wavefunction is $|\psi^{(2D)},--\rangle=\frac{1}{2}(1,-e^{-i\phi_x},-e^{-i\phi_y},e^{-i(\phi_x+\phi_y)})^T$, which is $|1/2,1/2,1/2,1/2\rangle$, the $s$ wave at $\Gamma$ ($k_x, k_y=0$) point. Similarly, the second and third energy bands have $s_1+s_2=0$, which are the waves $p_y$ ($|1/2,-1/2,1/2,-1/2\rangle$), and $p_x$ ($|1/2,1/2,-1/2,-1/2\rangle$) at the $\Gamma$ point. The fourth energy band has $s_1=1$ and $s_2=1$, the $d$ wave ($|1/2,-1/2,1/2,1/2\rangle$) at
the $\Gamma$ point~\cite{Obana2019,Cheng2022}. The bulk energy spectrum obtained from Eq.~(9) is plotted in Fig.~3(a).

\begin{figure}[t]
\leavevmode
\begin{center}
\leavevmode
\includegraphics[clip=true,width=0.99\columnwidth]{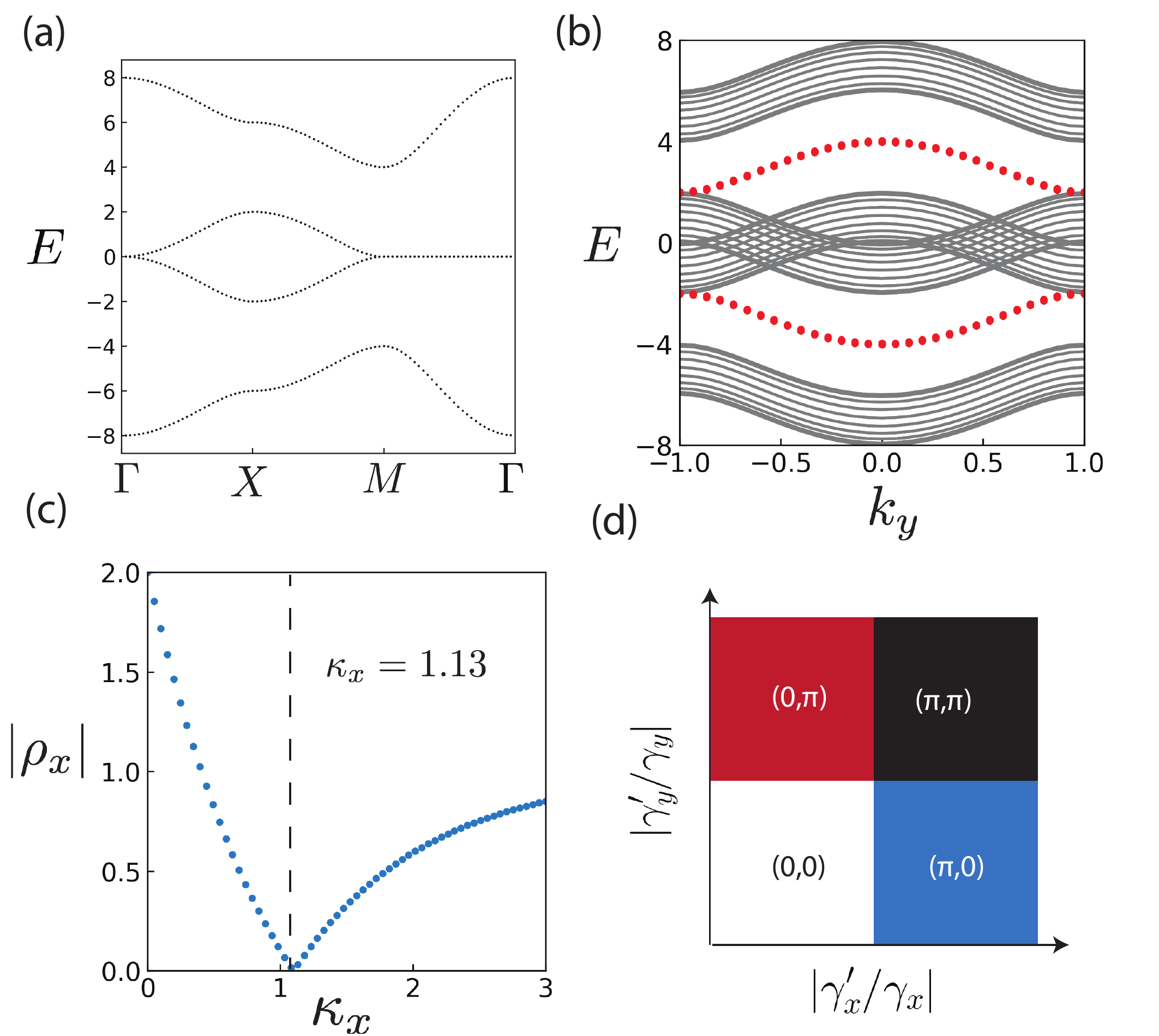}
\end{center}
\caption{ (a) Bulk energy spectrum of the 2D SSH model for $\gamma_x=\gamma_y=1.0$ and $\gamma^\prime_x=\gamma^\prime_y=3.0$.
(b) Ribbon spectrum of the 2D SSH model obtained by Eqs.~(9) and (10), where the gray lines are the bulk states and the red dots are the edge states. 
(c) $|\rho_x|$ dependence of $\pi+i\kappa_x$, where $|\rho_x|=0$ when $\kappa_x\approx 1.13$.
(d) Topological phase diagram of the 2D SSH model in terms of $\gamma^\prime_x/\gamma_x$ and $\gamma^\prime_y/\gamma_y$. }

\end{figure}

\subsubsection{Spectrum and wave functions of edge states}
In addition to the bulk energy spectrum, we can obtain the ribbon energy spectrum, including the edge states as well. For the ribbon structure, one direction is periodic and the other is finite. The energy spectrum of the 2D SSH model ribbon can be written as
\begin{equation}
    E^{2D}_x=s_1|\rho_x(K_x)|+s_2|\rho_y(k_y)|,
\end{equation}
where $k_y$ is the quasi-continuous wave number, $K_x$ is the discrete wave number that takes values of $-\pi/N, -\pi/N-1,\cdots \pi/N$ with $2N+1$ the number of the unit cells along the $x$-direction, and the subscript $x$ denotes that the ribbon is finite along the $x$-direction (we set lattice constant in all directions as unit).

For the edge states, $K_x=\pi+i\kappa_x$  with $\kappa_x$ the decaying length, which is determined by the ratio of $\gamma_x/\gamma_x^\prime$ together with the inversion symmetry as $\cosh\kappa=(\gamma_x^2+\gamma_x^{\prime2})/2\gamma_x\gamma_x^\prime$. 
It is noted that the parity of the edge states is determined by $s_1$. 
We plot Eq.~(10) for $\gamma/\gamma^\prime=1/3$ in Fig.~3(b), where the gray lines are the bulk energy bands and the red ones are the energy bands of the edge states. $\kappa_x$ is determined by $|\rho_x(\pi+i\kappa_x)|=0$. Figure 3(c) shows the dependence $|\rho_x|$ on $\kappa_x$. When $|\rho_x|=0$, the edge states are doubly degenerate, as required by the inversion symmetry.

The wavefunction of a finite sample of the 2D SSH model can be constructed using the linear combination of the bulk eigenstates with opposite wavenumber as formula as 
\begin{equation}
|v,s_1s_2,\mathbf{k}\rangle=C_{\mathbf{k}}|u,s_1s_2,\mathbf{k}\rangle+C_{-\mathbf{k}}|u,s_1s_2,-\mathbf{k}\rangle,    
\end{equation}
where $C_{\pm\mathbf{k}}$ has four components for each sublattice and the bulk eigenfunction $|u,s_1s_2,\mathbf{k}\rangle$ is 
\begin{widetext}

\begin{equation}
    |u,s_1s_2,\mathbf{k}\rangle=\sum^{N_x,N_y}_{1,1}
    e^{i(k_xm_x+k_ym_y)}
    \begin{pmatrix}
        1\\
        s_1e^{-i\phi_x}\\
        s_2e^{-i\phi_y}\\
        s_1s_2e^{-i(\phi_x+\phi_y)}
    \end{pmatrix}
    \begin{pmatrix}
        |(m_x,m_y),00\rangle, & |(m_x,m_y),01\rangle, & |(m_x,m_y),10\rangle, & |(m_x,m_y),11\rangle
    \end{pmatrix},
\end{equation}

\end{widetext}
where $(m_x,m_y)$ denotes the index of the unit cell. For the edge states that decay along the $x$-direction, the wavefunction is given as 
\begin{equation}
\begin{split}
|u^\text{edge},s_1s_2, k_y\rangle \sim
\sum^{N_x,N_y}_{m_x,m_y}
(-1)^{m_x}e^{-\kappa m_x}
\begin{pmatrix}
1\\
s_1\\
s_2e^{-i\phi_y}\\
s_1s_2e^{-i\phi_y}
\end{pmatrix}
\end{split},
\end{equation}
where a Bloch phase $e^{ik_ym_y}$ are omitted.

\subsubsection{Corner states}
The 2D SSH model can have topological corner states. For the corner states, they decay along both the $x$- and $y$- directions, whose wave number has the form $\mathbf{k}=(\pi+i\kappa_x,\pi+i\kappa_y)$, where $\kappa_i$ is the decaying length of the corner state along the $i$-direction. The extra phase $\pi$ is due to the open boundary condition, as discussed later. The energy of the corner states is given as 
\begin{equation}
\begin{split}
    E^{2D}_{x,y}=&s_1\sqrt{\gamma_x^2+\gamma_x^{\prime 2}-2\gamma_x\gamma_x^\prime\cosh{\kappa_x}}\\
    &+s_2\sqrt{\gamma_y^2+\gamma_y^{\prime 2}-2\gamma_y\gamma_y^\prime\cosh{\kappa_y}}.
\end{split}
\end{equation}
When the 2D SSH model has the $C_4$ point group symmetry, the four corner states differed by the values of $s_1$ and $s_2$ should be degenerate. In this case, $\rho_x(\pi+i\kappa_x)=\rho_y(\pi+i\kappa_y)=0$, and the corner states are fixed at zero energy. It is noted that for the 2D SSH model, we can define a chiral symmetry operator $\mathcal{C}=\sigma_z\otimes\sigma_z$ that $\mathcal{C}^\dagger \mathcal{H}_{2D}\mathcal{C}=-\mathcal{H}_{2D}$, which can also be considered as the sublattice symmetry. 

The wavefunction of the corner state is given as 
\begin{equation}
    |u^\text{corner},s_1s_2\rangle \sim 
    (1,s_1,s_2,s_1s_2)^T.
\end{equation}
where we see that the four corner states should appear as two pairs of opposite energies. Constrained by the chiral symmetry $\mathcal{C}$, these four corner states are degenerate at zero energy.

\subsubsection{Wilson loop and vectored Zak phase}
The Wilson loop, in general, can characterize the topological properties of quantum systems in terms of parallel transport. The 2D SSH model is not an exception either. In terms of quasi-momentum $\mathbf{k}$, the Wilson loop can be written as 
\begin{equation}
\Omega[C]=\mathcal{P}\exp(-i\oint_C d\mathbf{k} \cdot \mathbf{A}(\mathbf{k})),
\end{equation}
where $\mathcal{P}$ is path-ordering operator and $\mathbf{A}_{\alpha,\beta}=\langle \alpha, \mathbf{k}|i\nabla_\mathbf{k}|\beta,\mathbf{k}\rangle$ is Berry connection matrix. Taking the $k_x$-direction as an example and substituting the bulk wavefunction into Eq. (16), we have 
\begin{equation}
\begin{split}
\ln(\Omega_x)&=\int_{\Gamma}^{\Gamma=(2\pi/a,0)}A_xdk_x
\\&=
\begin{pmatrix}
\Delta \phi_x/2 &0 &-\Delta \phi_x/2 &0\\
0& \Delta \phi_x /2 & 0 &-\Delta \phi_x/ 2\\
-\Delta \phi_x/2 &0 & \Delta \phi_x/2 & 0\\
0&-\Delta \phi_x/2& 0 & \Delta \phi_x/2
\end{pmatrix},
\end{split}
\end{equation}
where $\Delta \phi_x=\phi_x(2\pi/a)-\phi(0)$. There are two eigenvalues of the Wilson loop matrix, which are $\Delta \phi_x$ and $0$. The different eigenvalues of the Wilson loop matrix can be considered as gauge choices. For the trivial eigenvalue 0, the corresponding eigenvectors are linear combinations of the $s$ and $p$ bands, which cannot distinguish the topology between the $|\gamma/\gamma^\prime|>1$ and $|\gamma/\gamma^\prime|<1$ regions, since the total winding of the two bands is always trivial. For the eigenvalue of $\Delta \phi_x$, the eigenvectors are each single band, respectively, and thus can be used to distinguish the topology between the two regions.

By choosing the single-band representation, the Wilson loop can be reduced to the Zak (Berry) phase. As there are two primary directions of the reciprocal space, the Zak phase in the 1D SSH model should be vectorized to distinguish the topology in all cases. The vectored Zak phase can be written as $\mathbf{Z}=(\mathcal{Z}_x,\mathcal{Z}_y)$. The emergence of topological edge and corner states in the 2D SSH model can be characterized by the vectored Zak phase, as $(\mathcal{Z}_x, \mathcal{Z}_y)$.  Figure 3(d) depicted the topological phase diagram of the 2D SSH model in terms of the vectored Zak phase depending on $|\gamma^\prime_i/\gamma|$.
For the vectored Zak phase $(\pi,\pi)$, there exist imaginary solutions of the wave number for both $x$- and $y$-directions, as discussed in Sec.~III.

\subsection{3D SSH model}

\subsubsection{Bulk spectrum and wave function}
For the 3D SSH model, its energy spectrum can be written as 
\begin{equation}
    E^{3D}=E^{2D}+s_3|\rho_z|,
\end{equation}
where $s_3=\pm 1$. The energy spectrum of the 3D SSH model is plotted in Fig.~3(a), where we have eight energy bands. The bulk wavefunction of the 3D SSH model is given as
\begin{equation}
|\psi^{3D},s_1s_2s_3\rangle=
\begin{pmatrix}
    1\\
    s_1e^{-i\phi_x}\\
    s_2e^{-i\phi_y}\\
    s_1s_2e^{-i(\phi_x+\phi_y)}\\
    s_3e^{-i\phi_x}\\
    s_1s_3e^{-i(\phi_x+\phi_z)}\\
    s_2s_3e^{-i(\phi_y+\phi_z)}\\
    s_1s_2s_3e^{-i(\phi_x+\phi_y+\phi_z)}\\
\end{pmatrix}.
\end{equation}
At the $\Gamma$ point, we have $\phi_i=\pi$, and the different values of $s_i$ correspond to the eight distinct eigenstates of the cubic point group.

\begin{figure}[t]
\leavevmode
\begin{center}
\includegraphics[clip=true,width=0.99\columnwidth]{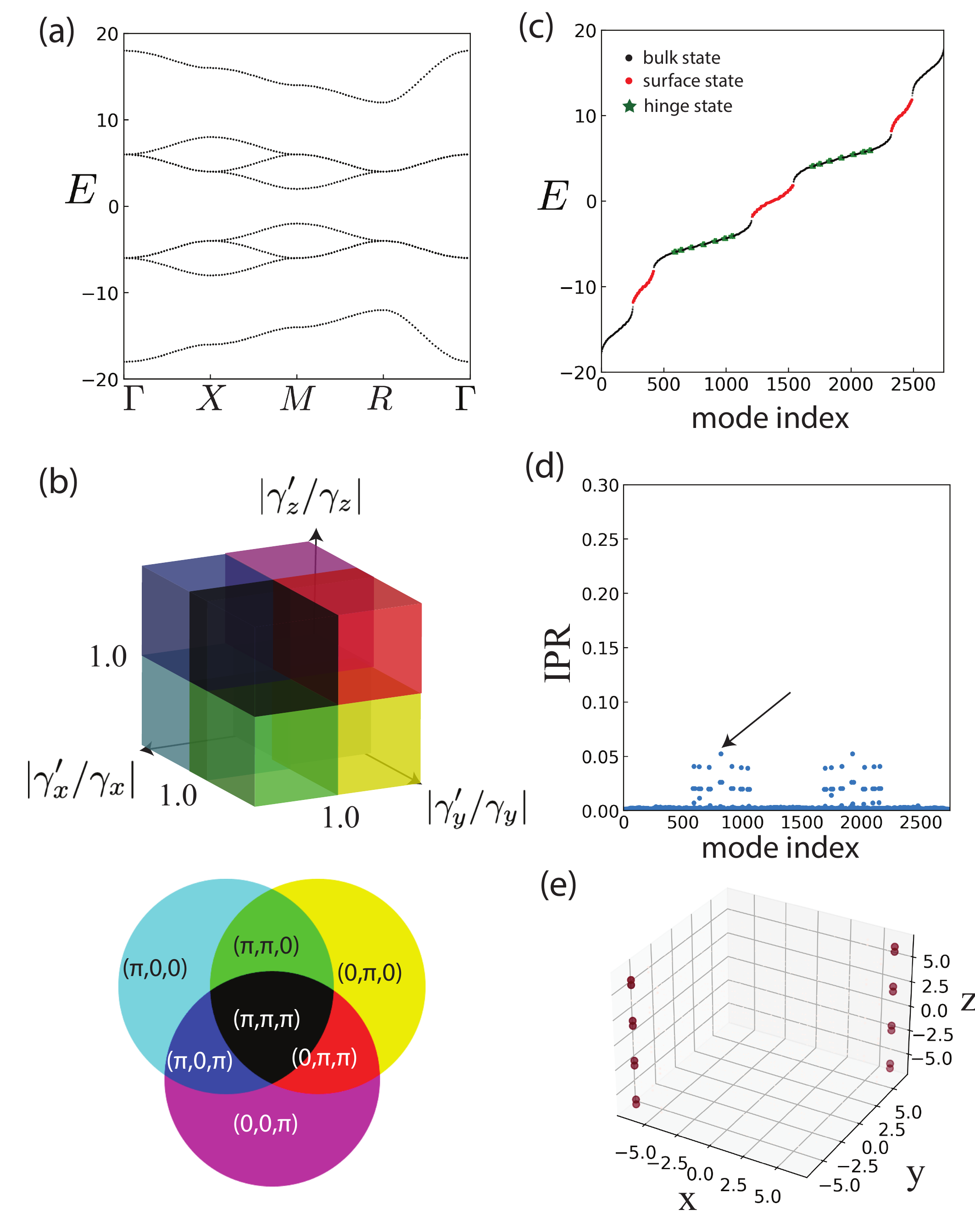}
\end{center}
\caption{(a) Bulk energy spectrum of the 3D SSH model along the high symmetric lines of the 1st Brillouin zone, which is obtained from Eq.~(16) for 0$\gamma_i=-1.0$ and $\gamma^\prime_i=-5.0$ with $i=x,y,z$. 
(b) Topological phase diagram of the 3D SSH model in terms of the ratio $|\gamma_i/\gamma_i^\prime|$ and the vectored Zak phase $\mathcal{Z}_i$.
(c) Eigenenergies for a finite sample of the 3D SSH model with $(\gamma_x,\gamma_y,\gamma_z)=(-1.0,-1.0,-5.0)$ and $(\gamma^\prime_x,\gamma^\prime_y,\gamma^\prime_z)=(-5.0,-5.0,-1.0)$. 
(d) Inverse participation ratio of the charge densities in (b), where the corner states are missing. 
(e) Charge density distribution of the intermediate-order topological hinge state as indicated by the arrow in (c). 
}
\end{figure}

\subsubsection{Intermediate-order topological phases}
Unlike the 2D SSH model, there is an intermediate-order topological phase in the 3D SSH model, which is neither lowest-order nor highest-order. 
In the 2D SSH model, its topological phases can be characterized by the vectored Zak phase, where $(\pi,\pi)$ indicates a topological phase of the highest order accompanied by the corner states, $(\pi, 0)$ is a topological phase of the lowest order accompanied by edge states along the $y$-direction, and $(0,0)$ is a trivial topological phase. 
For the 3D SSH model, except for the highest order topological phase of the vectored Zak phase $(\pi,\pi,\pi)$ and the lowest order topological phase of the vectored Zak phases $(\pi,0,0)$, $(0,\pi,0)$ and $(0,0,\pi)$, there are topological phases that are neither lowest order nor highest order like $(\pi,\pi, 0)$. 
We may call these topological phases intermediate-order topological phases, which are characterized by hinge states without corner states in 3D cases. 
As shown in Fig. 4(b), eight distinct topological phases are determined by $|\gamma_i/\gamma^\prime_i|$ for $i=x,y,z$, and characterized by the vectored Zak phase $(\mathcal{Z}_x,\mathcal{Z}_y,\mathcal{Z}_z)$ represented by different colors.

Taking the intermediate-order topological phase that has a vectored Zak phase $(\pi,\pi, 0)$ as an example, we expect that the second-order topological states, like the hinge states, appear, but the corner states are absent.
This can be verified by the numerical calculation of a finite sample of the 3D SSH model of $\gamma_x/\gamma_x^\prime=1/3$,$\gamma_y/\gamma_y^\prime=1/3$, and $\gamma_z/\gamma^\prime_z=3$. As shown in the eigenenergy spectrum of the finite 3D SSH model in Fig. 4(c), no zero-energy corner state appears for this parameter setting. Furthermore, by checking the inverse participation ratio of eigenstates as shown in Fig.~4(d), there are localized eigenstates. Graphing one of these localized eigenstates in Fig.~3(e), we find they are the expected hinge states. It is noted that these hinge states do not appear at zero energy due to the finite energy $|\rho_z|$.

Here, we discuss the topological protection of these intermediate-order states. Unlike the highest-order topological states like corner states, these lowest-order and intermediate-order states only reflect the band topology in the partial Brillouin zone. For example, the hinge states in the 3D SSH model are determined by the band inversions at $X$ and $Y$ points of the Brillouin only. In other words, no global gaps are required for the lowest-order and intermediate-order topological states. This fact adds extra robustness to these topological states, which we dub hierarchical topological phases, as discussed in detail later. These intermediate-order topological states are determined only by the band topology in the partial Brillouin zone, we call them fractional topological states. 

For the cases of $n>3$ that go beyond the physical dimensions, a possible realization is to use the synthetic dimensions. For example, using photonic modes of different frequencies in a photonic crystal~\cite{Yuan2018}.

\section{$n$-$(n-l)$ correspndence}

\subsection{ $n$D vectored Zak phase}
After discussing the 2D and 3D SSH models, we can extend the vectored Zak phase to arbitrary dimensions and apply it to the $n$D SSH model. 
Thus, each dimension can define its own winding number $\nu_i$ separately. 
The topological invariant of the $n$D SSH model can be a vector $\mathbf{Z}$ consisting of a series of winding numbers such as $(\nu_x,\nu_y,\nu_z,\cdots)$. 
Recalling that the Zak phase is nothing more than the Wannier center, it is clear that the topological invariant of the $n$D SSH is its Wannier center. 
Compared to the strong topological phase, such as the Haldane model, the topology of the $n$D SSH model is an atomic-obstructed phase. 
There is no obstruction in defining its Wannier center, and its topological edge states are due to the filling anomaly induced by the mismatch between the atomic and Wannier centers~\cite{Bradlyn2017}. The fractional topological phase has similar topological indices as weak topological insulators, and thus can be considered as a type of weak topology.

\begin{figure}[t]
\leavevmode
\begin{center}
\includegraphics[clip=true,width=0.99\columnwidth]{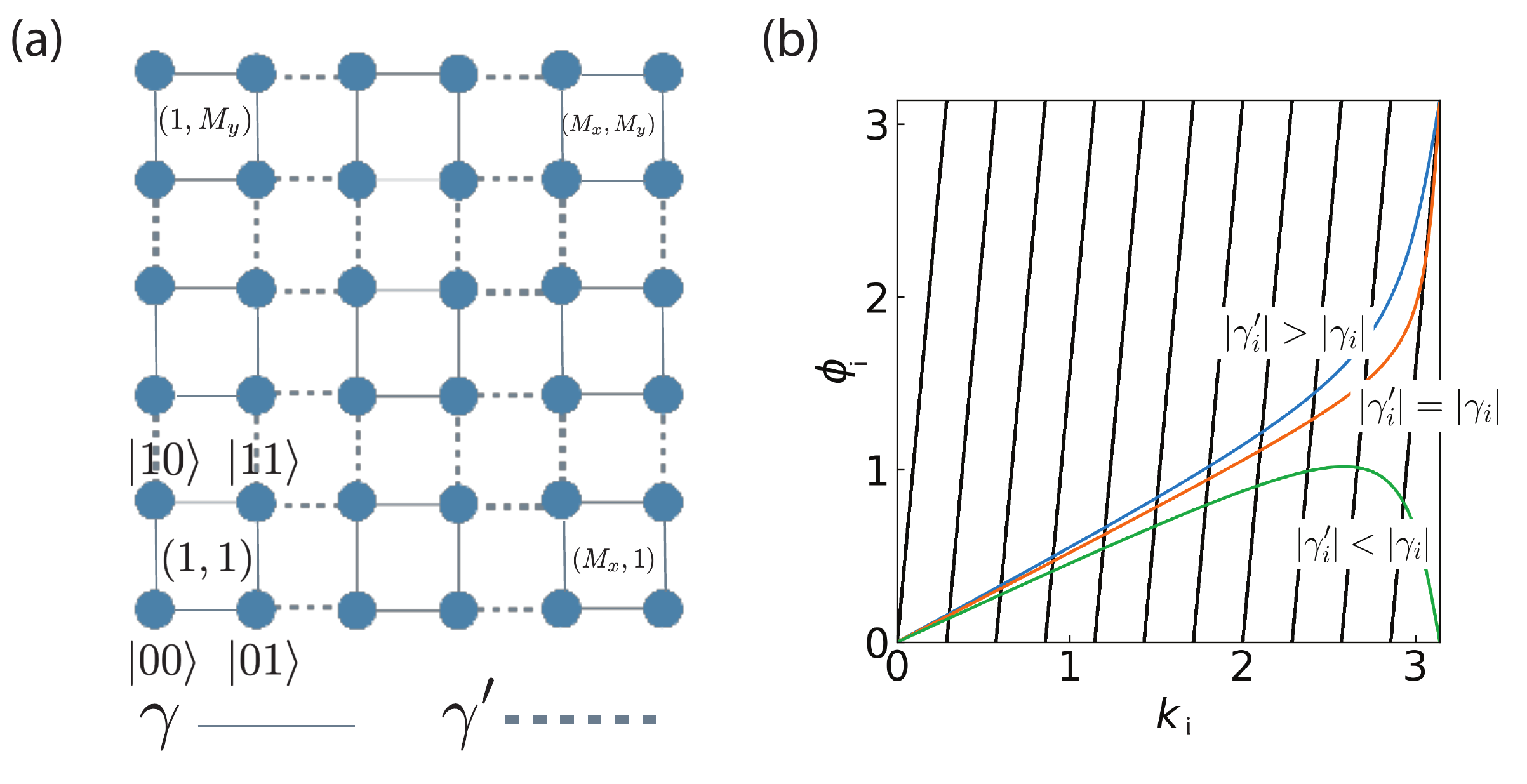}
\end{center}
\caption{(a) A finite sample of 2D SSH model with open boundaries condition.
There are four types of sublattice, labeled $|00\rangle$, $|01\rangle$, $|10\rangle$, and $|11\rangle$. 
The total number of unit cells is $M_x\times M_y$. 
(b) Graphing solving of the quantization condition Eq.~(19) for $M_i=10$. \
Depending on $|\gamma^\prime/\gamma|>1+1/(M+1)$ or not, there are $M-1$ and $M$ real solutions of the quasi-wavenumber $k_i$.}
\end{figure}

\subsection{Generalization of the bulk-edge correspondence}
Here, using the open boundary condition, we build the $n$-$(n-l)$ correspondence in the $n$D SSH model in terms of $\mathbf{Z}$. 
$\mathbf{Z}$ is a bulk topological invariant, which describes the topological in each direction by $\mathcal{Z}_i$. 
Taking the 2D SSH model as an example, we can connect the nontrivial $\mathcal{Z}_i$ and a purely imaginary solution of quasi-wavenumber $\kappa_i$ along the $i$-direction.  
Figure 4(a) illustrates a finite 2D SSH model with open boundary conditions, where a central index labels each unit cell and has four types of sublattices: $|00\rangle$, $|01\rangle$, $|10\rangle$, and $|11\rangle$. 
By imposing the open boundary condition along the $x$-direction~\cite{Delplace2011}, we obtain the following
\begin{equation}
    \begin{split}
    \langle(0,m_y),11|\nu,k_y=0\rangle&=0\\
    \langle(0,m_y),01|\nu,k_y=0\rangle&=0\\
    \langle(M_x+1,m_y),10|\nu,k_y=0\rangle&=0\\
    \langle(M_x+1,m_y),00|\nu,k_y=0\rangle&=0,\\
    \end{split}
\end{equation}
where $|\nu\rangle=C_\mathbf{k}|u_\mathbf{k}\rangle+C_\mathbf{-k}|u_\mathbf{-k}\rangle$ same as Eq.~(11). Taking Eqs.~(11) and (12) into Eq.~(18) and supposing the coefficients are the same for the sublattices in the same row, we obtain a quantization condition of $k_x$, which is 
\begin{equation}
    k_x(M_x+1)-\phi_x(k_x)=\tau_x\pi.
\end{equation}
Equation (21) is critical because it has $M_x$ or $M_x-1$ real roots, depending on the winding of $\phi_x(k_x)$.  As shown in Fig.~4(b), the lines set $f(k)=(M_x+1)k-\tau_x\pi$ with $\tau_x=1,2,\cdots,M_x$ coincide with $\phi_x(k_x)$ by $(M_x-1)$ times if $|\gamma_x^\prime/\gamma_x|>1+1/(M_x+1)$ with the extra term $1/(M_x+1)$ accounting for the finite size effect. According to the fundamental theorem of algebra, the missing solution around $\pi$ must be located in a complex regime, which is $k_x=\pi+i\kappa_x$. Thus, the nontrivial winding of $\phi_x$ directly leads to an imaginary solution of $k_x$, corresponding to a topological interfacial state.

The above discussion can also be generalized to any direction: a complex solution $k_i=\pi+i\kappa_i$ appears if $|\gamma^\prime_i/\gamma_i|>1$ for a large system size. 
We can define a $l$th higher-order topological invariant $\mathcal{Q}$ given by the product of $\mathcal{Z}_i$ as
\begin{equation}
\mathcal{Q}^{(l)}=\prod_i^l \mathcal{Z}_i,
\end{equation}
which builds the $n$-$(n-l)$ correspondence. Specifically, when all directions' $\mathcal{Z}_i$s are nontrivial, it results in $n$-0 correspondence, i.e., the appearance of 0D corner states. It is noted that this definition of the higher-order topological invariant by the product of Zak phases along distinct directions is applied to the $n$D SSH model only, and the generalization of $\mathcal{Q}^{(l)}$ to other systems is not trivial~\cite{Ono2019, Hwang2019}.

\subsection{Subsymmetry and hierarchy}

\begin{figure}[t]
\leavevmode
\begin{center}
\includegraphics[clip=true,width=0.99\columnwidth]{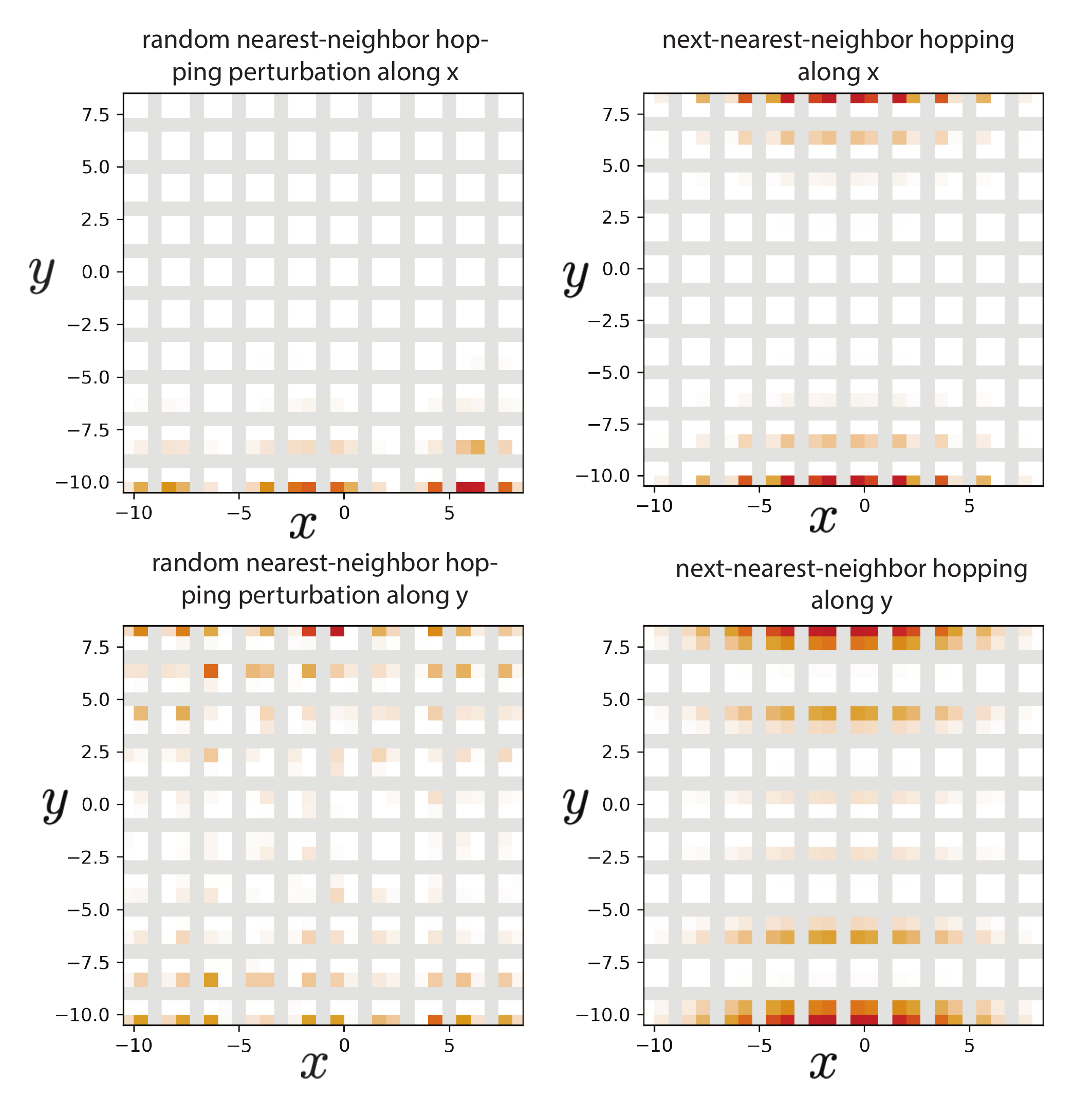}
\end{center}
\caption{Comparison of edge states along the $x$-direction under perturbations along different directions. The amplitudes of perturbations are $|\gamma^\prime|-|\gamma|$. The upper two panels are the perturbations along the $x$-direction, and the left one is the random perturbation on the nearest-neighbor hopping that keeps the chiral symmetry, while the right one is the perturbation that breaks the chiral symmetry like the next-nearest-neighbor hopping between the same type sublattices. The lower two panels are perturbations similar to the upper ones but along the $y$-direction.}
\end{figure}

Let's first define the chiral operator $\mathcal{C}$ for the $n$D SSH model, which is written as 
\begin{equation}
    \mathcal{C}=\prod^n_{i=1} \sigma_z,
\end{equation}
with the product the direct product. The topological states in the $n$D SSH model can always be paired with opposite eigenenergies according to Eq.~(4). Constrained by $\mathcal{C}$, all those topological $0$D states are degenerate and thus are bound to zero energy. It is noted that the chiral symmetry can be further released to the subsymmetry, where not all $0$D states are bound to zero energy, but a partial of them as $\mathcal{C}$ can be decomposed into a series of sublattice symmetries~\cite{Milad2022, Wang2023}.  

In addition to the highest-order 0D topological states, the first- and intermediate-order topological states do not require complete chiral symmetry $\mathcal{C}$ to maintain their robustness. This is due to the unique hierarchical structure of the $n$D SSH model as indicated by Eq.~(3). We can define a subchiral symmetry operator as 
\begin{equation}
\mathcal{C}^\prime=\prod_i^l\sigma_z,
\end{equation}
where $l$ corresponds to the $l$th-order topological states, and $\sigma_z$ is picked up within the $2^n$ sublattice space. 

Taking the edge states along the $x$-direction of the 2D SSH model as an example, they are impervious to hopping perturbations in the $x$-direction even if they break the complete chiral symmetry $\mathcal{C}$. As displayed in Fig.~6, the edge states are resilient to hopping perturbations along the $x$ direction, even for the next-nearest-neighbor hopping connecting the same sublattices, which breaks the chiral symmetry. This fact implies the importance of the hierarchical structure in the $n$D SSH model protecting the topological states. It is noted that even for the 1D SSH model, it can be considered as the piling up of two 0D SSH models (that is, two single sites connecting by $\rho$), and the chiral symmetry can be decomposed into two sublattice symmetries. Thus, we can dub the SSH-like topological insulators as hierarchical topological insulators, where a hierarchical relation exists between its $n$D and $(n-1)$D versions, and their topological interfacial states are protected by a subsymmetry related to the hierarchical structure. In particular, hierarchical topological insulators are characterized by intermediate-order topological interfacial states, like the hinge states in the 3D SSH model, which only reflect the band topology in the partial Brillouin zone. 


\section{Comparison with the BBH model}

\begin{figure}[t]

\begin{center}
\includegraphics[clip=true,width=0.99\columnwidth]{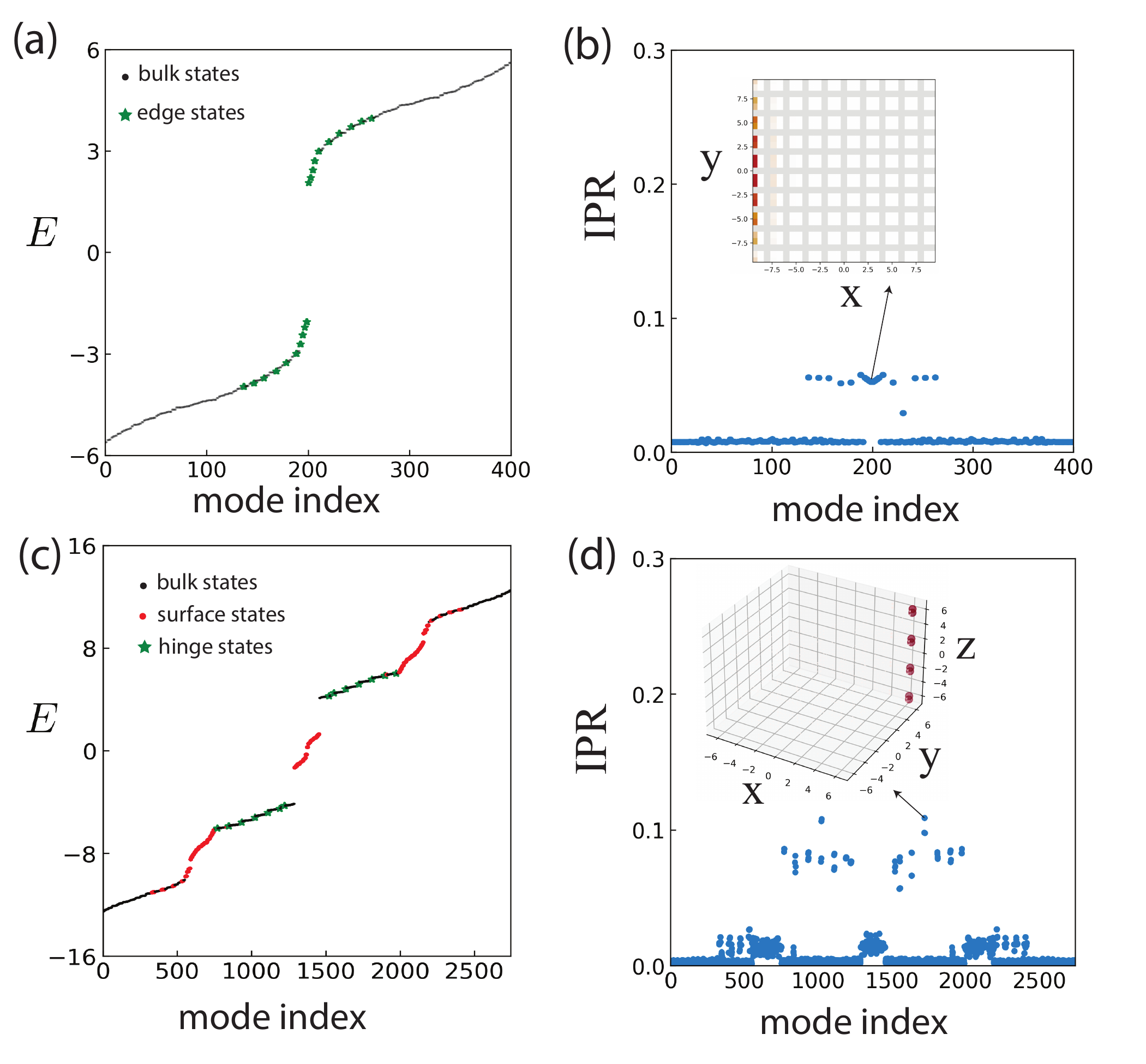}
\end{center}
\caption{(a) Eigenenergy spectrum of the 2D BBH model for $\gamma_x=\gamma^\prime_y=1.0$ and $\gamma_y=\gamma^\prime_x=3.0$. 
(b) IPR of the 2D BBH model in (a). The inset shows the charge density distribution of the topological edge state. Corner states are missing.
(c) Eigenenergy spectrum of the 3D BBH model for $(\gamma_x,\gamma_y,\gamma_z)=(1.0,1.0,5.0)$ and $(\gamma^\prime_x,\gamma^\prime_y,\gamma^\prime_z)=(5.0,5.0,1.0)$
(d) Inverse participation ratio of the 3D BBH model in (c). Inset is the charge density distribution of the eigenstate with the largest IPR.}
\end{figure}

\subsection{Spectrum and wave function of bulk}
Compared to the 2D SSH model, the 2D BBH model has a flux of $\pi$ in each unit cell. 
By choosing one of the $y$-direction hoppings as the carrier of the $\pi$ flux phase, we can write the BBH model as 
\begin{equation}
H^\text{BBH}_{2D}=
\begin{pmatrix}
    0 & \rho_x & -\rho_y & 0 \\
    \rho^*_x & 0 & 0 & \rho_y \\
    -\rho^*_y & 0 & 0 & \rho_x \\
    0 & \rho_y & \rho^*_x &0 
    
\end{pmatrix},
\end{equation}
where the bases are the sublattice bases of the 2D SSH model such as $|00\rangle$, $|01\rangle$, $|10\rangle$, and $|11\rangle$. 
Written in a similar form to the Eq.~(5), we have 
\begin{equation}
H^\text{BBH}_{2D}=
\begin{pmatrix}
    H^\text{BBH}_{1D} & \rho_y\sigma_z\\
    \rho^*_y\sigma_z & H^\text{BBH}_{1D}
\end{pmatrix},
\end{equation}
where$H^\text{BBH}_{1D}$ is the Hamiltonian of the 1D BBH model that is same as the 1D SSH model. By rearranging the sublattice bases that ensure the diagonal block is zero, Eq.~(25) can be rewritten as 
\begin{equation}
H^\text{BBH}_{2D}=
    \begin{pmatrix}
        0&0&-\rho_y&\rho_x\\
        0&0&\rho^*_x&\rho^*_y\\
        -\rho^*_y&\rho_x&0&0\\
        \rho^*_x & \rho_y&0&0        
    \end{pmatrix}.
\end{equation}
Using the determinant of the block matrix $\det\begin{pmatrix}
    A & B \\
    C & D
\end{pmatrix}=\det(AB-CD)$ if $C$ and $D$ commute, we have the eigenvalue $(E^2-|\rho_x|^2-|\rho_y|^2)^2=0$. The doubly degenerate energy spectrum of the 2D BBH model is 
\begin{equation}
    E_{\pm}=\pm \sqrt{|\rho_x|^2+|\rho_y|^2},
\end{equation}
and the corresponding eigenfunctions are
\begin{equation}
    |\psi^\text{BBH}_{2D},\pm\rangle=
    \begin{pmatrix}
        -\frac{\rho_y}{E_\pm}\\
        1\\
        0\\
        \frac{\rho^*_x}{E_\pm}
    \end{pmatrix}
    \text{,  }
    |\psi^{\prime \text{BBH}}_{2D},\pm\rangle=
    \begin{pmatrix}
        \frac{\rho_x}{E_\pm}\\
        0\\
        1\\
        \frac{\rho^*_y}{E_\pm}
    \end{pmatrix}
    ,
\end{equation}
where the bases are $(|00\rangle, |01\rangle, |10\rangle, |11\rangle)$, and  the normalization factor $1/\sqrt{2}$ is omitted. Unlike the 2D SSH model, the eigenstates in the 2D BBH model are doubly degenerate.

Similarly, we can obtain the Hamiltonian of the 3D BBH model, which is written as
\begin{equation}
H^\text{BBH}_{3D}=
\begin{pmatrix}
H^\text{BBH}_{2D} & \rho_z\sigma_z\\
\rho^*_z\sigma_z & H^\text{BBH}_{2D},
\end{pmatrix}
\end{equation}
where the sublattice bases are $|000\rangle$, $|011\rangle$, $|010\rangle$, $|001\rangle$,  and $|100\rangle$, $|111\rangle$, $|110\rangle$, $|101\rangle$. 
After rearranging the sublattices, the 3D BBH Hamiltonian can be cast into an off-diagonal form as  
\begin{equation}
H^\text{BBH}_{3D}=
\begin{pmatrix}
    0&0&0&0&-\rho_y&\rho_x&-\rho_z&0\\
    0&0&0&0&\rho^*_x&\rho^*_y&0&-\rho_z\\
    0&0&0&0&0&\rho^*_z&\rho^*_x&\rho_y\\
    0&0&0&0&\rho^*_z&0&-\rho^*_y&\rho_x\\
    -\rho^*_y&\rho_x&0&\rho_z&0&0&0&0\\
    \rho^*_x&\rho_y&\rho_z&0&0&0&0&0\\
    -\rho^*_z&0&\rho_x&-\rho_y&0&0&0&0\\
    0&-\rho^*_z&\rho^*_y&\rho^*_x&0&0&0&0
\end{pmatrix}
\end{equation}
with eigenvalues $(E^2-|\rho_x|^2-|\rho_y|^2-|\rho_z|^2)^4=0$. The eigenstates can then be solved accordingly, and only half of the sublattices are independent.

\subsection{Edge and hinge states without corner states}
Imposing the same open boundary condition as the 2D SSH model to the 2D BBH model, we obtain a similar quantization condition of $k_x$, which is 
\begin{equation}
    k_x(M_x+1)-\phi_x(k_x)=\tau_x\pi .
\end{equation}
This fact suggests that in the 2D BBH model, the topological edge state can exist alone, similar to the 2D SSH model. 
Figures 7(a) and (c) show the energy spectrum and the charge density distribution of the topological edge states for a finite sample of the 2D BBH model with $|\gamma_x|<|\gamma^\prime_x|$
and $\gamma_y|>|\gamma^\prime_y|$. 
As expected, there is no zero-energy corner state for this parameter setting.

Furthermore, the BBH model can host intermediate-order states as well. 
As displayed in Figs.~7(c) and (d), 
for parameter setting: $(\gamma_x,\gamma_y,\gamma_z)=(1.0,1.0,5.0)$ and $(\gamma^\prime_x,\gamma^\prime_y,\gamma^\prime_z)=(5.0,5.0,1.0)$, 
there are topological hinge states, but no corner states. 
The emergence of intermediate-order topological states in the BBH model suggests that the BBH model is another example of the hierarchical topological insulator, similar to the $n$D SSH model.

\subsection{Topological invariant}
The Wilson loop can also characterize the topological property of the BBH model. Taking the 2D case as an example and setting $\gamma_i=0$ for simplicity, the Wilson loop matrix of the BBH model along the $k_y$-direction is given as 
\begin{equation}
\ln(\Omega_y)=
\begin{pmatrix}
      \frac{\cos^2\theta\Delta \phi_y}{2}  &- \frac{\cos^2\theta\Delta \phi_y}{2} &0 & 0\\
      - \frac{\cos^2\theta\Delta \phi_y}{2}  &  \frac{\cos^2\theta\Delta \phi_y}{2} &0 & 0\\
      0&0 &  -\frac{\cos^2\theta \Delta \phi}{2} &\frac{\cos^2\theta \Delta \phi}{2} \\
     0&0 &\frac{\cos^2\theta\Delta \phi_y}{2}  &  -\frac{\cos^2\theta \Delta \phi_y}{2}
\end{pmatrix},
\end{equation}
where $\cos\theta=|\rho_y|/E_+$ and $\sin\theta=|\rho_x|/E_+$. Solving the above Wilson loop matrix, we obtain the eigenvalues 0 and $\pm \cos^2\theta \Delta \phi_y$. For the nontrivial $\pm \cos^2\theta \Delta \phi_y$, it is also determined by the winding of $\phi_i$, similar to the $n$D SSH model. For finite $\gamma_i$, it shares the same topology as the $\gamma_i=0$ case due to continuous change through adiabatic evolution. The modulation function of $\Delta \phi_y$ for finite $\gamma_i$ can be written approximately as $(\cos^2\theta-\theta^2\cos2\theta/2)$.

\subsection{$n$D BBH model}
The 2D and 3D BBH models can be generalized to the $n$D case following a hierarchical structure similar to that of the $n$D SSH model. By choosing a proper ordering of the sublattice bases that ensures every dimension carries a $\pi$ flux, the $n$D Hamiltonian of the BBH model can be written as
\begin{equation}
H^\text{BBH}_{nD}=
\begin{pmatrix}
    H^\text{BBH}_{(n-1)D} & \rho_i\sigma_z\\
    \rho^*_i\sigma_z & H^\text{BBH}_{(n-1)D}
\end{pmatrix},
\end{equation}
where $\sigma_z$ is the Pauli matrix, and $H^\text{BBH}_{(n-1)D}$ is the Hamiltonian of $(n-1)$D BBH model. 
The eigenvalues of the $n$D BBH model are $E=\pm\sqrt{\sum_i|\rho_i|^2}$~\cite{Luo2023A, Luo2023B}. The bulk-wave function can then be solved similarly to the 2D and 3D BBH models.

\section{Summary}

In summary, we have observed a hierarchical structure in the $n$D SSH model and obtained analytical solutions for the $n$D SSH model. These solutions include a quantization condition for the quasi-wavenumber, leading to interfacial states for nontrivial winding numbers. We have generalized the bulk-edge correspondence from the 1D SSH model to an $n$-$(n-l)$ correspondence for arbitrary dimensions, revealing a new type of topological insulator known as a hierarchical topological insulator. This hierarchical topological insulator can host intermediate-order topological states that reflect band topology only in a partial Brillouin zone.
Furthermore, we compared the BBH and $n$D SSH models and discovered that they share a similar hierarchical structure, and the BBH model is another example of a hierarchical topological insulator. 


 \section*{Acknowledgments}
This work is supported by the Ningbo University Research Starting Funding, NSFC Grant No. 12074205, and NSFZP Grant No. LQ21A040004. F. Liu appreciates the useful discussion with Ce Shang and the constructive advice of the referees.


%

\end{document}